**Algebraic approach to exact solution of the (2 + 1)-dimensional Dirac oscillator**

**in the noncommutative phase space**

H. Panahi[1] and A. Savadi[2]

*Department of Physics, University of Guilan, Rasht 41635-1914, Iran*

**Abstract**

In this paper we study the (2 + 1)-dimensional Dirac oscillator in the noncommutative phase space and the energy eigenvalues and the corresponding wave functions of the system are obtained through the sl(2) algebraization. It is shown that the results are in good agreement with those obtained previously via a different method.

*Keywords*: Dirac oscillator, noncommutative phase space, sl(2) algebraization.

PACS number: 03.65.-w, 03.65.Pm, 03.65.Fd, 03.65.Ge

## 1. Introduction

Study on Dirac oscillator as an important potential has attracted a lot of attention and has found many physical applications in various branches of physics [1-6]. The Dirac oscillator was introduced for the first time by Ito et al [7], in which, the momentum $\vec{p}$ in Dirac equation is replaced by $\vec{p} - im_0 w \beta \vec{r}$, where $\vec{r}$ is the position vector and $m_0, w$ and $\beta$ are the mass of particle, the frequency of the oscillator and the usual Dirac matrices respectively. Similar system was studied by Moshinsky and Szczepaniak [8], who gave it the name of Dirac oscillator, due to in the non-relativistic limit, it becomes a simple harmonic oscillator with strong spin-orbit coupling term. In quantum optics and for (2+1)-dimension space, it is seen that the Dirac oscillator system can be mapped into Anti-Jaynes-Cummings model [9-11] which describes the atomic transitions in a two level system. In Refs. [5, 12], it has been shown that the Dirac oscillator interaction can be interpreted as the interaction of the anomalous magnetic moment with a linear electric field. Also the electromagnetic potential associated with the Dirac oscillator interaction has been found by Benitez et al in Ref. [13].

---

[1] Corresponding author E-mail: t-panahi@guilan.ac.ir

[2] E-mail: a.savadi@gmail.com

On the other hands in recent decades, the noncommutativity has become an extremely active area of research such as in string theories, quantum field theories and in quantum mechanics [14-24]. For quantum systems in noncommutative space, it is seen that assuming the noncommutativity may be a result of quantum gravity effects, also it is a fruitful theoretical laboratory where we can get some insight on the consequences of noncommutativity in field theory by using standard calculation techniques of quantum mechanics. The study on Dirac equation in noncommutative phase space has also attracted much attention in recent years [25-31]. For example, study on the relativistic Landau levels of Dirac equation in (2+1)- dimensional noncommutative phase space has been shown in Ref. [28] and one can see an exact mapping of this relativistic model to the AJC model. Also in Ref. [25], it is seen that for Landau problem in two dimensional noncommutative phase space, the equation of motion of a harmonic oscillator is similar to the equation of motion of a particle in a constant magnetic field and in the lowest Landau level. The energy gap of Dirac oscillator in a noncommutative phase space changes by noncommutative effect [25]. A noncommutative description of graphene, which consists of a Dirac equation for massless Dirac fermions plus noncommutative corrections has been studied in Ref. [27] and has been shown that the momentum noncommutativity affects the energy levels of grapheme.

Since the appearance of quantum mechanics, considerable efforts have been devoted to obtaining exact solutions of the relativistic and non-relativistic wave equations, using different methods and techniques [32-36]. Now in this paper, we study the Dirac oscillator in noncommutative phase space within the framework of representation theory of the sl(2) Lie algebra.

A quantum system is exactly solvable (ES) if all the eigenvalues and the corresponding eigenfunctions can be calculated in an exact analytical manner. In contrast, a quantum system is quasi-exactly solvable (QES) if only a finite number of eigenvalues and eigenfunctions can be determined exactly [37-40]. In the case of ES models, the Hamiltonian of the system can be diagonalized completely due to the fact that algebraic symmetry is complete, but in the case of QES models, the Hamiltonian is only block diagonalized. Such a finite block can always be diagonalized, which yields a finite part of the spectrum algebraically.

This paper is organized as follows. In section 2, based on Ref. [31], we briefly review the noncommutative phase space on $(2 + 1)$-dimensional Dirac oscillator. Section 3 is devoted to $sl(2)$ algebraization of Dirac oscillator in the noncommutative phase space. We obtain the energy eigenvalues and the corresponding wave functions through the sl(2) representation and show that our results are in good agreement with the results of Ref. [31]. In section 4, we present the conclusion.

## 2. Review on $(2 + 1)$-dimensional Dirac oscillator in the noncommutative phase space

According to Ref. [31], the noncommutative phase space is characterized by the fact that their coordinate operators satisfy the equation

$$\left[ x_i^{(NC)}, x_j^{(NC)} \right] = i \tilde{q}_{ij}^{\%}, \quad \left[ p_i^{(NC)}, p_j^{(NC)} \right] = i \overline{q}_{ij}, \quad \left[ x_i^{(NC)}, p_j^{(NC)} \right] = i \mathbf{h} d_{ij}, \tag{1}$$

where $\tilde{q}_{ij}^{\%}$ and $\overline{q}_{ij}$ are an antisymmetric tensor of space dimension. By using the Bopp's shift method, the two dimensional noncommutative phase space can be considered as [29]

$$x^{(NC)} = x - \frac{\tilde{q}^{\%}}{2\mathbf{h}} p_y, \quad y^{(NC)} = y + \frac{\tilde{q}^{\%}}{2\mathbf{h}} p_x, \quad p_x^{(NC)} = p_x + \frac{\overline{q}}{2\mathbf{h}} y, \quad p_y^{(NC)} = p_y - \frac{\overline{q}}{2\mathbf{h}} x. \tag{2}$$

The Dirac oscillator is also described by the following Hamiltonian [8]

$$\hat{H} = c\hat{a}\cdot\left(\vec{p} - im_0 w \beta \vec{r}\right) + \beta m_0 c^2, \tag{3}$$

where in two-dimensional noncommutative space, it is transformed as

$$\left\{ca_x\left(p_x^{(NC)} - im_0 w\beta x^{(NC)}\right) + ca_y\left(p_y^{(NC)} - im_0 w\beta y^{(NC)}\right) + \beta m_0 c^2\right\}y_D = E_{NC}y_D. \tag{4}$$

By considering the Pauli matrices as the representation of Dirac matrices in (2+1)-dimension, that is

$$a_x = s_x = \begin{pmatrix} 0 & 1 \\ 1 & 0 \end{pmatrix}, \quad a_y = s_y = \begin{pmatrix} 0 & -i \\ i & 0 \end{pmatrix}, \quad \beta = \begin{pmatrix} 1 & 0 \\ 0 & -1 \end{pmatrix}, \tag{5}$$

also $y_D = \begin{pmatrix} y_1 & y_2 \end{pmatrix}^T$, then Eq. (4) becomes

$$\begin{pmatrix} m_0 c^2 & cp_- \\ cp_+ & -m_0 c^2 \end{pmatrix}\begin{pmatrix} y_1 \\ y_2 \end{pmatrix} = E_{NC}\begin{pmatrix} y_1 \\ y_2 \end{pmatrix}, \tag{6}$$

where $p_-$, $p_+$ are as follows:

$$p_- = p_x^{(NC)} - ip_y^{(NC)} + im_0 w\left(x^{(NC)} - iy^{(NC)}\right) = r_1\left(p_x - ip_y\right) + im_0 w r_2\left(x - iy\right), \tag{7}$$

$$p_+ = p_x^{(NC)} + ip_y^{(NC)} - im_0 w\left(x^{(NC)} + iy^{(NC)}\right) = r_1\left(p_x + ip_y\right) - im_0 w r_2\left(x + iy\right), \tag{8}$$

with

$$r_1 = 1 + \frac{m_0 w}{2\hbar}\bar{q}, \quad r_2 = 1 + \frac{\bar{q}}{2m_0 w\hbar}. \tag{9}$$

After some calculations, it is easy to see that the equation (6) gives

$$\left\{c^2 p_- p_+ - \left(E_{NC}^2 - m_0^2 c^4\right)\right\}y_1 = 0, \tag{10}$$

$$\left\{c^2 p_+ p_- - \left(E_{NC}^2 - m_0^2 c^4\right)\right\}y_2 = 0. \tag{11}$$

Similar to Refs. [31, 41] and by defining $p_x = p\cos q$, $p_y = p\sin q$, $p^2 = p_x^2 + p_y^2$, it is seen that the equations (7) and (8) transform into

$$p_- = e^{-iq}\left\{r_1 p - I\left(\frac{\partial}{\partial p} - \frac{i}{p}\frac{\partial}{\partial q}\right)\right\}, \tag{12}$$

$$p_+ = e^{iq}\left\{r_1 p + I\left(\frac{\partial}{\partial p} + \frac{i}{p}\frac{\partial}{\partial q}\right)\right\}, \tag{13}$$

where

$$l = \left(1 + \frac{\bar{q}}{2m_0 w \mathbf{h}}\right) m_0 w \mathbf{h}. \tag{14}$$

Substituting Eqs. (12) and (13) into (10) and by considering $y_1(p,q) = f(p)e^{imq}$, we have

$$\left(\frac{d^2 f(p)}{dp^2} + \frac{1}{p}\frac{df(p)}{dp} - \frac{m^2}{p^2}f(p)\right) + \left(k^2 - k'^2 p^2\right)f(p) = 0, \tag{15}$$

where

$$k^2 = \frac{2l\,r_1(m+1)+e}{l^2}, \qquad k'^2 = \frac{r_1^2}{l^2}, \tag{16}$$

with

$$e = \frac{E_{NC}^2 - m_0^2 c^4}{c^2}. \tag{17}$$

Here it is necessary to point out that in the literature, there are many different methods for solving quantum models exactly [32-36]. Boumali and Hassanabadi [31] solved this problem analytically and obtained the exact solutions in terms of the hypergeometric functions. Within the present study, we intend to illustrate the exact solvability of the problem through the $sl(2)$ algebraization [39]. To do a comprehensive analysis, we need to obtain the more general solution of the problem. This is the main advantage of the present method. In fact, by using the theory of representation space, we construct the general matrix equation of the problem and calculate the closed-form expressions for energies and eigenfunctions. Accordingly, we can quickly obtain exact solutions of any arbitrary state *n,* without any cumbersome numerical procedure or a complicated analytical one in determining the solutions of the higher states. It seems that the method is computationally much simpler than other analytical methods.

Hence in the next section, we solve the equation (15), by Lie algebraic approach related to representation theory of sl(2) Lie group. Similar procedure can be also done on Eq. (11) for the $y_2$ spinor wave function.

## 3. Lie algebraic solution of the Dirac oscillator in the noncommutative phase space

In this section, for solving the equation (15) by the sl(2) Lie algebra representation, we first use similarity transformation as $f(p) = p^{-1/2} R(p)$, such that the first order derivative of the differential equation (15) can be eliminated, and it is reduced to a Schrodinger-type operator as

$$\left( \frac{d^2}{dp^2} + \frac{1/4 - m^2}{p^2} - k^2 - k^2 p^2 \right) R(p) = 0. \tag{18}$$

Extracting the asymptotic behavior of the wavefunction at the origin and infinity, using the transformation

$$R(p) = p^g e^{ap^2} F(p), \tag{19}$$

where $a < 0$ (to be physically meaningful ) and $g$ are parameters to be determined later by the exact solvability conditions (Eq. (28)), Eq. (18) is converted to

$$\frac{d^2 F(p)}{dp^2} + \left( \frac{2g}{p} + 4ap \right) \frac{dF(p)}{dp} + \left( \left( 4a^2 - k^2 \right) p^2 + \frac{g(g-1) + 1/4 - m^2}{p^2} + 4ag + 2a + k^2 \right) F(p) = 0. \tag{20}$$

Using the change of variable $r = kp^2$ we get

$$r \frac{d^2 F}{dr^2} + \left( \frac{2a}{k} r + g + \frac{1}{2} \right) \frac{dF}{dr} + \left[ \left( \frac{a^2}{k^2} - \frac{1}{4} \right) r + \frac{g(g-1) + 1/4 - m^2}{4r} + \frac{ag}{k} + \frac{a}{2k} + \frac{k^2}{4k} \right] F = 0. \tag{21}$$

According to Ref. [38], in one dimension, the only Lie algebra of the first-order differential operators, which possesses finite-dimensional representations, is the sl(2) Lie algebra. Hence we consider the following realization of the operators

$$J_n^+ = r^2 \frac{d}{dr} - nr, \qquad J_n^0 = r \frac{d}{dr} - \frac{n}{2}, \qquad J_n^- = \frac{d}{dr}, \tag{22}$$

which satisfy the sl(2) commutation relations

$$\left[ J_n^+, J_n^- \right] = -2 J_n^0, \qquad \left[ J_n^\pm, J_n^0 \right] = \mathbf{m} J_n^\pm, \tag{23}$$

and preserve the (n + 1)-dimensional linear space of polynomials with finite order

$$P_{n+1} = \left\langle 1, r, r^2, ..., r^n \right\rangle. \tag{24}$$

The most general second-order differential equation which preserves the space $P_{n+1}$, can be alsp represented as a quadratic combination of the sl(2) generators as

$$H = \sum_{a,b=0,\pm} C_{ab} J_n^a J_n^b + \sum_{a=0,\pm} C_a J_n^a + C,$$  (25)

where $C_{ab}, C_a$ and $C$ are real parameters. Substitution (22) into (25), yields the following differential form:

$$H = P_4(r)\frac{d^2}{dr^2} + P_3(r)\frac{d}{dr} + P_2(r),$$  (26)

where $P_i(r)$ are polynomials of degree $i$

$$P_4(r) = C_{++}r^4 + C_{+0}r^3 + C_{+-}r^2 + C_{0-}r + C_{--},$$

$$P_3(r) = C_{++}(2-2n)r^3 + \left(C_+ + C_{+0}\left(1-\frac{3n}{2}\right)\right)r^2 + (C_0 - nC_{+-})r + \left(C_- - \frac{n}{2}C_{0-}\right),$$  (27)

$$P_2(r) = C_{++}n(n-1)r^2 + \left(\frac{n^2}{2}C_{+0} - nC_+\right)r + \left(C - \frac{n}{2}C_0\right).$$

Comparing Eq. (26) with (21) gives

$$C_{++} = C_{+0} = C_{+-} = C_{--} = C_+ = 0,$$

$$a = -\frac{k}{2}, g = m + \frac{1}{2},$$  (28)

$$C_{0-} = 1, \ C_0 = -1, \ C_- = \frac{n}{2} + m + 1, \ C = \frac{k^2}{4k} - \frac{1}{2}(m+n+1).$$

From Eqs. (25) and (28), the Lie algebraic form of the Hamiltonian is written as

$$H = J_n^0 J_n^- - J_n^0 + \left(\frac{n}{2} + m + 1\right)J_n^- + \left(\frac{k^2}{4k} - \frac{1}{2}(m+n+1)\right),$$  (29)

which implies that this operator is ES, due to the absence of the terms of positive gradings. Therefore, the operator H preserves the finite-dimensional representation space of the algebra sl(2) as

$$F(r) = \sum_{j=0}^{n} a_j r^j, \qquad n = 0,1,2,\mathbf{K} \ .$$  (30)

Hence, using equations (29), (30) and doing some calculations, the following matrix equation can be obtained

$$
\begin{pmatrix}
\dfrac{k^2}{4k}-\dfrac{1}{2}(m+1) & (m+1) & 0 & \mathbf{L} & 0 \\[2mm]
0 & \dfrac{k^2}{4k}-\dfrac{1}{2}(m+1)-1 & 2+2(m+1) & \mathbf{L} & 0 \\[2mm]
0 & 0 & \mathbf{O} & \mathbf{O} & \mathbf{M} \\[2mm]
\mathbf{M} & \mathbf{M} & 0 & \dfrac{k^2}{4k}-\dfrac{1}{2}(m+1)-(n-1) & n(n-1)+n(m+1) \\[2mm]
0 & 0 & \mathbf{L} & 0 & \dfrac{k^2}{4k}-\dfrac{1}{2}(m+1)-n
\end{pmatrix}
\begin{pmatrix}
a_0 \\ a_1 \\ \cdot \\ \cdot \\ \cdot \\ a_{n-1} \\ a_n
\end{pmatrix} = 0, \quad (31)
$$

where the necessary condition for the closed form of the energy eigenvalues is as follows

$$
\frac{k^2}{4k}-\frac{1}{2}(m+1)-n=0, \qquad\qquad n=0,1,2,\mathbf{K}\ . \qquad (32)
$$

By substituting the values of $k^2$ and $k$ into (32) we have

$$
E_{NC_n}=\pm m_0 c^2 \sqrt{1+4n\left(1+\frac{m_0 w}{2\mathbf{h}}\widetilde{q}\right)\left(1+\frac{\overline{q}}{2m_0 w\mathbf{h}}\right)\frac{w\mathbf{h}}{m_0 c^2}}, \qquad (33)
$$

which is the same as the energy relation in Ref. [31].

Also from Eq. (31) the expansion coefficients $a_m$'s satisfy the following two-term recursion relation

$$
\left(j\left(j+1\right)+\left(j+1\right)(m+1)\right)a_{j+1}+\left(\frac{k^2}{4k}-\frac{1}{2}(m+1)-j\right)a_j=0, \qquad (34)
$$

with the boundary conditions $a_{-1}=0$ and $a_{n+1}=0$.

Before we proceed further, it would be well to note that the equation (33) may or may not possess real eigenvalues depending on the choice of noncommutative parameters $\widetilde{q}$ and $\overline{q}$ to be positive or negative values. Of course, it is not our aim to study here, but one can see a similar study in Ref. [42], where the Dirac oscillator in 2+1 dimensional noncommutative space has been studied for the cases of positive and negative noncommutative parameter .Hence it seems that the results can be used to study the connection with non-Hermitian quantum mechanics. Also, the possible case of imaginary energy gives useful clue regarding the lack of bound-states in the spectroscopy of relativistic fermions.

Therefore for a given $n$, the energy eigenvalue is obtained from (33) and the corresponding unnormalized wave function can be obtained from (6) as

$$Y_D \equiv y_{n,m}(p,q) = \begin{pmatrix} y_1(p,q) \\ y_2(p,q) \end{pmatrix} = \begin{pmatrix} 1 \\ \dfrac{cp_+}{E+m_0c^2} \end{pmatrix} y_1(p,q),$$ (35)

where

$$y_1(p,q) = p^m e^{-\frac{k}{2}p^2} e^{imq} \sum_{j=0}^{n} a_j r^j,$$ (36)

with $r = kp^2$. Now for clarifying that the wave function obtained from this method, is in good agreement with the results of Ref. [31], we calculate it for $n = 2$. According to Eqs.(30) and (34), the three dimensional invariant subspace is obtained as

$$F(r) = a_0 + a_1 r + a_2 r^2$$
$$= a_0 \left[ 1 + \frac{\left( \dfrac{-k^2}{4k} + \dfrac{1}{2}(m+1) \right)}{m+1}(kp^2) + \frac{\left( -\dfrac{k^2}{4k} + \dfrac{1}{2}(m+1) \right)\left( -\dfrac{k^2}{4k} + \dfrac{1}{2}(m+1)+1 \right)}{(m+1)\big(2+2(m+1)\big)}(kp^2)^2 \right], \quad (37)$$

where by using the definition of the confluent hypergeometric function for $n = 2$ we have

$$F(r) = a_0 \,_1F_1\left( -n, m+1; kp^2 \right).$$ (38)

The above equation is exactly identical to $Eq.(32)$ in Ref. [31] and so for other values of n and after some calculations, one can show that $F(r)$ can be written in terms of the confluent hypergeometric function.

# 4. Conclusions

Using the Lie algebraic approach, we have solved the Dirac oscillator in (2+1) dimension in the framework of noncommutative phase space. We have obtained the energy eigenvalues and the corresponding wave functions through the sl(2) Lie algebra representation. We have also shown that our results are in good agreement with the results in Ref. [31]. It is seen that the Lie algebraic approach is a powerful method for reproducing the exact analytical results.

## References

[1] A. Bermudez, M. A. Martin- Delgado and A. Luis, Phys. Rev. A **77**, 063815 (2008).

[2] N. Ferkous and A. Bounames, Phys. Lett A **325**, 21 (2004).

[3] Victor M. Villalba, Phys. Rev. A **49**, 586 (1994) .

[4] A. Bermudez, M. A. Martin-Delgado and E.Solano, Phys. Rev. A 76, 041801 (2007).

[5] M. Moreno and A. Zentella, J. Phys. A: Math. Gen **22**,  L821 (1989) .

[6] R. de Lima Rodrigues, " On the Dirac oscillator," Physics Letters A, vol. 372, no. 5, pp. 2587-2591, 2008.

[7] D. Ito, K. Mori and E. Carriere, Nuovo Cimento A **51**, 1119 (1967).

[8] M. Moshinsky and A. Szczepaniak, J. Phys. A **22**, L817 (1989).

[9] E. T. Jaynes and F. W. Cummings, Proc. IEEE **51**, 89 (1963).

[10] J. Larso, Phys. Scr. **76**, 146 (2007).

[11] L. Allen and J. H. Eberly, *Optical Resonance and Two Level Atoms, Dover Publications*, Mineola,New York, (1987).

[12] R. P. Martinez-y-Romero and A. L. Salas-Brito, J. Math. Phys, **33** , 1831 (1992).

[13]  J. Bernitez, R.P. Martinez, Y. Romero, H.N. Yepez and A.L. Salas-Brito, Phys. Rev. Lett **64**, 1643 (1990).

[14]   N. Seiberg and E. Witten, JHEP **09**, 032 (1999).

[15]  A. Connes, M. Douglas and A. S. Schwarz, JHEP **02**, 003 (1998).

[16]  M. R. Douglas and N. A. Nekrasov, Rev. Mod. Phys. **73**, 977 (2001).

[17]  C. S. Chu and P. M. Ho, Nucl. Phys. B **550**, 151 (1999).

[18]  C. S. Chu and P. M. Ho, Nucl. Phys. B **568**, 447 (2000).

[19]  F. Ardalan, H. Arfaei and M. M. Sheikh-Jabbari, Nucl. Phys. B **576**, 578 (2000).

[20] J. Jing and Z-W Long, Phys. Rev. D **72**, 126002 (2005).

[21] I. Hinchliffe, N. Kersting and Y. L. Ma, Int. J. Mod. Phys. A **19**, 179 (2004).

[22] S. Minwalla, M. Van Raamsdonk and N. seiberg, JHEP **02**, 020 (2000).

[23] M. Van Raamsdonk and N Seiberg, JHEP **03**, 035 (2000).

[24] R. Gopakumar, S. Minwalla and A.Strominger, JHEP **05**, 020 (2000).

[25] S. Cai, T. Jing, G. Guo and R. Zhang, Int. J. Theor. Phys. **49**, 1699 (2010).

[26] O. Bertolami and R. Queiroz, Phys. Lett. A **375**, 4116 (2011).

[27] Z. Y. Luo, Q. Wang, X. Li and J. Jing, Int. J. Theor. Phys. **51**, 2143 (2012).

[28] B. P. Mandal and S. K. Rai, Phys. Lett. A **376**, 2467 (2012).

[29] C. Bastos, O. Bertolami, N. C. Dias and J. N. Prata, Int. J. Mod. Phys. A **28**, 1350064 (2013).

[30] O. Panella and P. Roy, Phys. Rev. A **90**, 042111 (2014).

[31] A. Boumali and H. Hassanabadi, Zeitschrift für Naturforschung. A **70**, 619 (2015).

[32] H. Ciftci, R.L. Hall and N. Saad, J. Phys. A **36**, 11807 (2003).

[33] S.H. Dong, D. Morales and J. Garcia-Ravelo, Int. J. Mod. Phys. E **16**, 189 (2007).

[34] G.F. Wei and S.H. Dong, Can. J. Phys. **89**, 1225 (2011).

[35] H. Hassanabadi, S. Zarrinkamar and A. A. Rajabi, Commun. Theor. Phys. **55**, 541 (2011).

[36] H. Hassanabadi, A. A. Rajabi and S. Zarrinkamar, Mod. Phys. Lett. A **27**, 1250057 (2012).

[37] A.G. Ushveridze, Mod. Phys. Lett. A **5**, 1891 (1990).

[38] A.V. Turbiner, Commun. Math. Phys. **118**, 467 (1988).

[39] A.V. Turbiner, "Lie-algebras and linear operators with invariant subspaces," Contemporary Mathematics, vol. 160, pp. 263-310, 1994.

[40] D. Gomez-Ullate, N. Kamran and R. Milson, Phys. Atom. Nucl. **70**, 520 (2007).

[41] L. Menculini, O. Panella and P. Roy, Phys. Rev. D 87, 065017 (2013).

[42] Y. Q. Luo,· Y. Cui · Z. W. Long and J. Jing, Int. J. Theor. Phys. 50, 2992 (2011).